\documentclass[utf8]{article} 

\usepackage{url,hyperref,lineno,microtype,subcaption}
\usepackage[onehalfspacing]{setspace}
\usepackage{fullpage}
\usepackage{graphicx}

\title{Investigating charge trapping in ferroelectric thin films through transient measurements}

\author{Suzanne Lancaster\,$^{1,*}$, \and Patrick D. Lomenzo\,$^{1}$, \and Moritz Engl\,$^{1}$, \and Bohan Xu\,$^{1}$, \and Thomas Mikolajick\,$^{1,2}$, \and Uwe Schroeder\,$^{1}$ \and Stefan Slesazeck\,$^{1}$}

\date{
	$^1$NaMLab gGmbH, Germany \\ \texttt{suzanne.lancaster@namlab.com}\\%
	$^2$IHM, TU Dresden, Germany \\%
}

\begin{document}
\onecolumn
\maketitle

\begin{abstract}

A measurement technique is presented to quantify the polarization loss in ferroelectric thin films as a function of delay time during the first 100s after switching. This technique can be used to investigate charge trapping in ferroelectric thin films by analyzing the magnitude and rate of polarization loss. Exemplary measurements have been performed on Hf$_{0.5}$Zr$_{0.5}$O$_2$ (HZO) and HZO/Al$_2$O$_3$ films, as a function of pulse width and temperature. It is found that the competing effects of the depolarization field, internal bias field and charge trapping lead to a characteristic Gaussian dependence of the rate of polarization loss on the delay time. From this, a charge trapping and screening model could be identified which describes the dynamics of polarization loss on short timescales.

\end{abstract}

\section{Introduction}
Charge trapping at the interfaces of ferroelectric (FE) HfO$_2$-based thin films plays an important role in both the dynamics of polarization switching and FE device operation. Charge trapping has a significant impact on device operation in the case of FE field-effect transistors (FeFETs) \cite{ma2002nonvolatile,ali2018silicon} as well as ferroelectric tunnel junctions (FTJs) \cite{fontanini2022interplay}. In both of these devices, an intentional dielectric (DE) layer is added to a metal-ferroelectric-metal (MFM) stack, which will modulate the impact of charge trapping due to the additional FE-DE interface. Assuming fixed and trapped charge densities is necessary in order to fully model the switching behavior of ferroelectric devices with DE layers \cite{fontanini2021polarization, hoffmann2022intrinsic}. Nonetheless, charge trapping is also involved in the switching of HfO$_2$ ferroelectrics without any intentional interfacial layer (IL) \cite{mehta1973depolarization, islamov2019identification}, due to the unavoidable presence of non-switching 'dead layers' at the FE-electrode interfaces \cite{stengel2006origin}. It has been shown that even after switching, the injection of charges can continue to modify switching behaviour by shifting the coercive voltage through a so-called 'fluid imprint' effect \cite{buragohain2019fluid}. As such, a thorough understanding of charge trapping, including its origin, how it can be characterized and how it can be improved or harnessed, is clearly necessary, in order to optimize the operation of FE films in real devices. 

When a non-switching DE layer is introduced between a FE layer and an electrode, numerous impacts on the FE switching need to be considered. Non-switching layers generate a depolarization field which destabilizes the polarization state, leading to polarization loss \cite{lomenzo2019ferroelectric}. Even in the common case when ferroelectric HfO$_2$ is sandwiched between two TiN electrodes with no intentionally grown dielectric layers, TiO$_x$ and TiON can form at the ferroelectric-metal interfaces which will impact the depolarization field and interface trap states \cite{hamouda2020physical}. In the presence of a DE layer, a significant leakage current \cite{si2019ferroelectric} and high density of interface traps \cite{fontanini2021polarization} need to be considered to fully compensate charges arising from polarization switching. The degree to which polarization charges are compensated can be controlled via the switching pulse parameters, and thereby the charge injected during switching \cite{park2021polarizing}.

The impact of charge trapping on device operation has been quite heavily researched for FeFETs, where an interlayer improves the memory window for a ferroelectric integrated on a semiconductor channel \cite{mulaosmanovic2021ferroelectric}. In this case, trapped charges most notably cause threshold voltage instabilities \cite{yurchuk2016charge} which increases read latency, i.e. the time after writing at which the device can first be reliably read \cite{ni2018critical}. Finally, trapping and detrapping has been suggested as a driving force for gradual or accumulative switching in ferroelectric films, where domains are switched by the application of multiple consecutive pulses \cite{lancaster2022multi, mulaosmanovic2018accumulative}. 

In this paper we will investigate charge trapping in single-layer Hf$_{0.5}$Zr$_{0.5}$O$_2$ (HZO) and bilayer HZO/Al$_2$O$_3$ films with TiN electrodes. We present a novel measurement technique making use of a modified positive-up, negative-down (PUND) pulse train to quantify the polarization loss that occurs on short time scales (\textless30\,s) after switching the FE. This is a complementary method to standard retention measurements, which measure the polarization state after long time periods \cite{mueller2012reliability}. On the contrary, this method allows the quantification and analysis of fast polarization loss in ferroelectric films. This fast polarization loss is a consequence of competing internal fields and charge trapping effects in the time frame during which the film reaches a state of quasi-equilibrium. We will discuss how various parameters in the measurement such as temperature and pulse width can be modified in order to investigate the dynamics of polarization stabilization and loss, by presenting results performed on three film stacks with different top interfacial layers. A strongly varying rate of polarization loss is observed in the first 100\,ms after switching, which is a critical time period for ferroelectric device operation. Finally, further applications of the presented method are discussed.

\section{Materials and methods}
\subsection{Device fabrication}
In order to fabricate capacitor devices, blanket bottom electrodes (BEs) of TiN were first deposited via sputtering under ultra-high vacuum. Then, 11\,nm thick Hf$_{0.5}$Zr$_{0.5}$O$_2$ (HZO) films were deposited via atomic layer deposition (ALD) by alternating cycles of HfO$_2$ and ZrO$_2$ with HyALD (HfCp(NMe$_2$)$_3$) and ZyALD (ZrCp(NMe$_2$)$_3$) as metal-organic precursors and ozone as an oxidant. Al$_2$O$_3$ layers of different thicknesses (1.5 or 2\,nm) were deposited with ALD using TMA as a precursor and ozone as an oxidant. Top electrodes (TEs) of TiN were deposited in the same way as the BEs and the sample was then annealed at 500°C for 20s to promote crystallization of the HZO films. Finally, capacitor structures with a diameter of 200\,$\mu$m were formed by evaporating top contacts of Ti/Pt (10/25\,nm) through a shadow mask, which acted as hard masks for a subsequent SC1 etching of the top TiN electrodes. 

\subsection{Pulse train for quantifying backswitched polarization}\label{pulsetrain}

\begin{figure}[h!]
\begin{center}
\includegraphics[width=18cm]{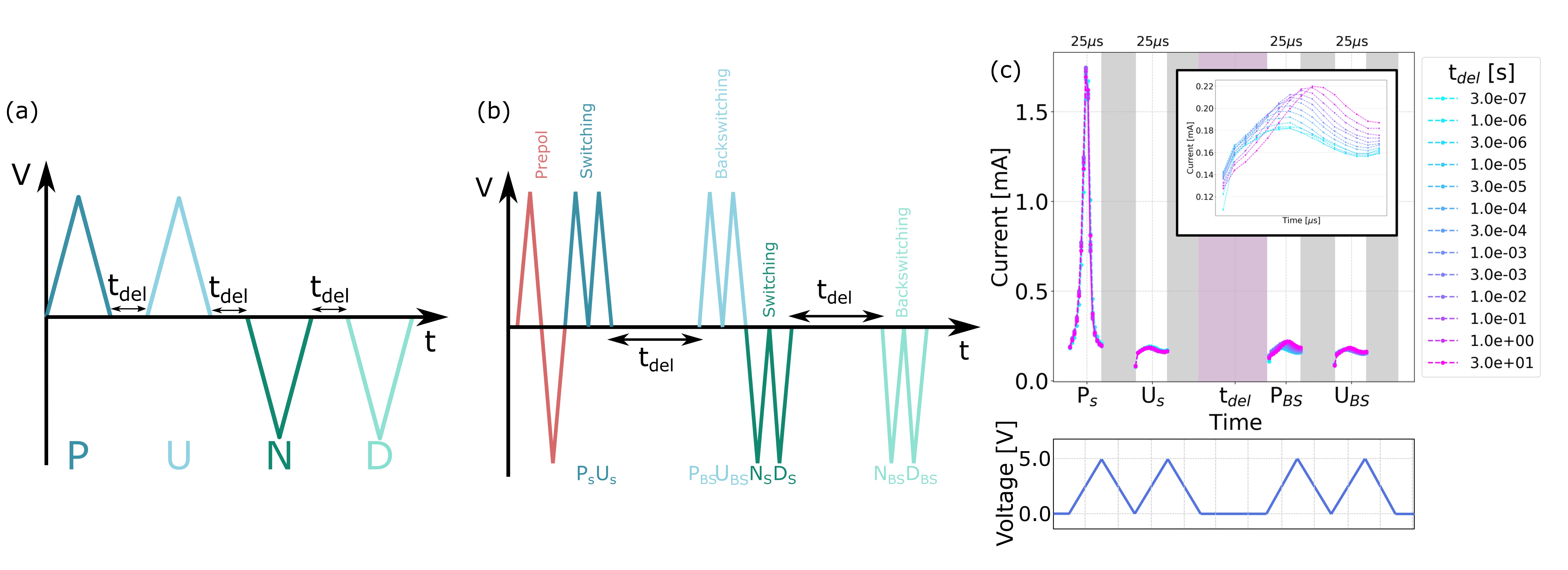}
\end{center}
\caption{(a) Standard PUND (Positive Up, Negative Down) pulse train for characterization of ferroelectric films;(b) pulse train for measuring polarization loss in ferroelectric thin films; (c) example of the current trace recorded when measuring the (P$_{loss}$) at different delay times. The backswitched current pulse in positive polarity (P$_{BS}$) is shown on a smaller scale in the inset.}\label{fig:1}
\end{figure}

To investigate polarization loss (P$_{loss}$) as a function of different pulse parameters, the pulse train shown in figure \ref{fig:1} was applied to the devices described above. Before the measurement sequence, a prepolarization pulse consisting of a triangular positive and negative pulse is applied to minimize any impact of imprint effects on the measurement results. The measurement sequence is a modified version of a PUND (Positive-Up, Negative-Down, figure \ref{fig:1}a) pulse sequence used to isolate the switching current in ferroelectric films \cite{Fukunaga_JPSJ2008}. Two sets of pulses are applied in each polarity, separated by a time\ delay t$_{del}$ at 0\,V, where the first pulse pair determines the switching current, and the second again switches any polarization which was lost during t$_{del}$. In this case, double pulses are applied for each switching or backswitching measurement, where the first pulse contains all current contributions from switching, leakage, and dielectric displacement (\textit{P/N} pulses), while the second is performed immediately after the first pulse, so that only leakage and dielectric displacement currents contribute (\textit{U/D} pulses). The current difference between these pulses can therefore be integrated over time to find the switched charge. The different pulses in each polarity are denoted with \textit{S} for the switching pulses and \textit{BS} for the backswitching pulses. When changing polarity and in between the pre-pole pulses, a time delay of 500\,ns with an applied voltage of 0\,V is used to minimize the influence on the subsequent pulse. 

An example of the current traces measured for all time delays for a single set of pulse parameters (pulse width and amplitude), at a fixed temperature, is shown in figure \ref{fig:1}b. Due to measurement constraints, the switching charge is always integrated only on the rising edge of the pulse. In preliminary investigations, it was found that the current measured on the falling edge is always identical in the switching and non-switching pulses. As such, this contribution always cancels completely in the calculation of the switched charge. This is shown with exemplary measurements in the supplementary materials (figure S1). 

P$_{loss}$ can then be quantified via the following equation:
\begin{equation}\label{eqn:1}
P_{loss} = \int_{0}^{t_{pls}/2} (P_{BS} - U_{BS}) \,dt
\end{equation}
and may be normalized to the total switched charge by:
\begin{equation}\label{eqn:2}
\tilde{P_{loss}} = 100\cdot \frac{\int_{0}^{t_{pls}/2}(P_{BS} - U_{BS})\,dt }{\int_{0}^{t_{pls}/2}( P_{S} - U_{S})\,dt }
\end{equation}
In the negative polarity, the \textit{P} and \textit{U} pulses are replaced with \textit{N} and \textit{D} pulses. 

The inset in \ref{fig:1}b shows the current of $P_{BS}$ on an enlarged scale. It is clear that with an increasing $t_{del}$, the coercive voltage $V_c$ of the $P_{BS}$ shifts to higher voltages, an effect known as imprint. The physical meaning of this shift is that as $t_{del}$ increases, it becomes energetically more difficult to switch all domains back into the same state. This phenomenon has been described as a signature of charge trapping into the interfacial layers during the delay time \cite{tagantsev2006interface}. The imprint effect provides valuable information on the redistribution of charges during the delay time and could further be investigated for its time-dependence \cite{chernikova2021dynamic, takada2021time}. In this paper however, we focus primarily on quantifying the polarization loss as an integral of the switching current.  

\subsection{Measurement parameter space}
\begin{figure}[h!]
\begin{center}
\includegraphics[width=17cm]{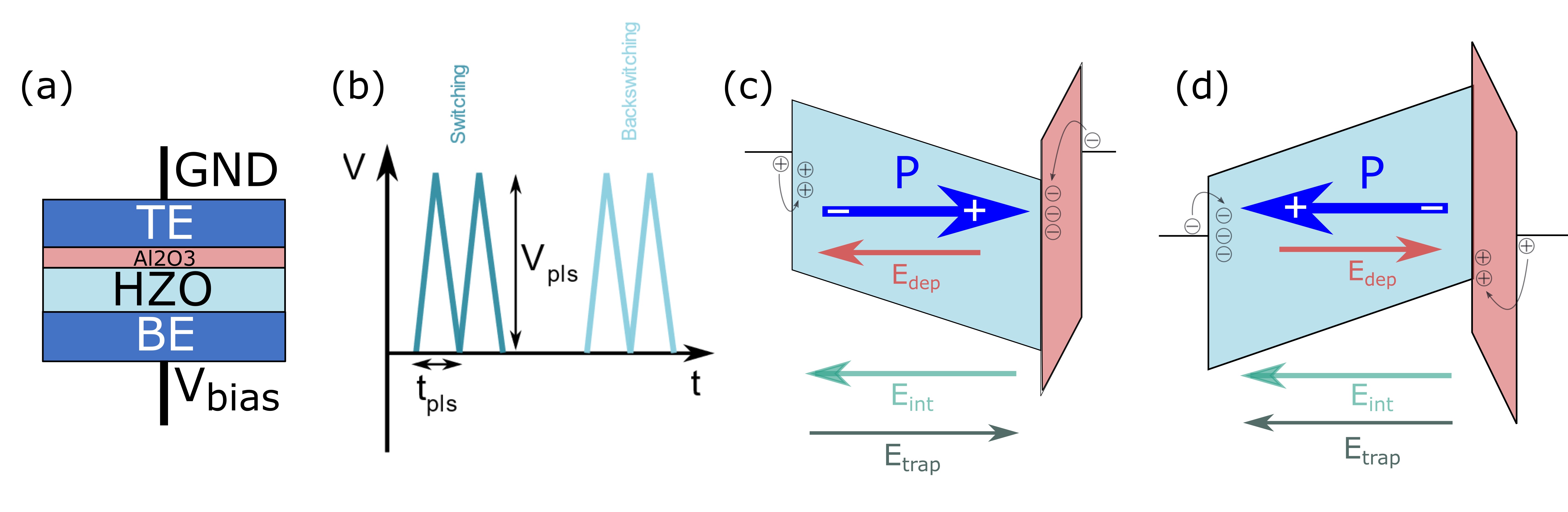}
\end{center}
\caption{(a) Stack schematic of a HZO/Al$_2$O$_3$ bilayer with the ground and bias indicated; (b) applied voltages as a function of time, indicating pulse width (varied in our experiments) and height; band diagrams of the HZO/Al$_2$O$_3$ device stack in the (c) P$_{up}$ state (positive polarity) and (d) P$_{down}$ state (negative polarity)}\label{fig:2}
\end{figure} 

Figure \ref{fig:2}a depicts the capacitor stack with the direction of applied voltage. A positive bias would facilitate electron injection at the TE during switching, while negative bias encourages electron injection at the BE. Likewise, a positive bias can inject holes at the BE, while a negative bias can inject holes at the TE, although electron injection is expected to be a stronger effect \cite{alyabyevaferroelectric}. In order to investigate the impact of charge trapping, various pulse parameters are modified. Modulation of the internal electric field by charge injection is achieved by changing the pulse width t$_{pls}$ as depicted in figure \ref{fig:2}b, which is kept the same for all pulses on the pulse train. The pulse widths applied in our experiment were 25, 50, 100 and 250\,$\mu$s. The peak voltage V$_{pls}$ is fixed so that the field over the FE layer (assuming a capacitive voltage divider model and neglecting the influence of any unintentional interlayers, discussed below) is $\sim$\,3.5\,V. One caveat is that the range of t$_{pls}$ possible to apply for a given V$_{pls}$ is somewhat limited, since full switching on all pulses is required so as not to distort the analysis. 

Here it should be noted that besides the intentional dielectric layer at the top electrode, previous experiments have demonstrated that due to the deposition process, a TiO$_2$/TiON layer is formed at the bottom electrode interface \cite{baumgarten2021impact}. Without an Al$_2$O$_3$ layer at the top of the HZO film, TiON also forms at the TE interface with HZO, which may further grow during high temperature annealing \cite{hamouda2020physical}. While these have a smaller influence on the depolarization field and field drop over the ferroelectric, they cannot be neglected and impact the switching of the HZO film due to the poor uniformity and chemical heterogeneity of these interfaces \cite{hamouda2020physical, baumgarten2021impact}, which have a significant impact on the ferroelectric behavior of MFM HZO capacitors with TiN electrodes.

The simplified band diagrams for the HZO/Al$_2$O$_3$ bilayers under no external applied bias are shown schematically in figures \ref{fig:2}c \& \ref{fig:2}d for P$_{up}$ and P$_{down}$, respectively. The field inside the ferroelectric is made up of several components, as shown on the diagram. Opposing the polarization there is always a depolarization field E$_{dep}$, whose magnitude depends on the thickness of the non-switching DE layer, the relative dielectric constants of the DE and FE, and the magnitude of the polarization, P \cite{lomenzo2020depolarization}. In addition, an internal bias field E$_{int}$ exists due to work function differences in the electrodes \cite{pesic2018built} and/or to fixed charges in the ferroelectric, namely oxygen vacancies formed at the TE during annealing \cite{fengler2018relationship}. In all of our stacks, E$_{int}$ points from the top to the bottom electrode, i.e. it points in the same direction as the depolarization field when the device is in the P$_{up}$ state.

Countering this and thus stabilizing the ferroelectric polarization by providing an enhanced screening charge, trapped charges in the dielectric layer produce an opposing field E$_{trap}$ \cite{fontanini2022interplay, alyabyevaferroelectric}. Besides carrier injection during switching, it is expected that additional electrons and holes can be trapped during the delay time t$_{del}$, driven by the depolarization field and/or the internal field (as indicated in the diagrams). Thus during t$_{del}$ the system should tend towards an equilibrium, with polarization loss occurring and charge trapping continuing until reaching a point where all the internal fields are sufficiently weak to cause unobservable changes at the device terminals. The rate of polarization loss would then depend on the interplay of the static internal bias field, the depolarization field, and the screening field caused by injected carriers.

Since E$_{dep}$ is dependent on the properties of the dielectric layer, and furthermore the thicknesses of the unintentional layers cannot be determined with high accuracy, we will investigate three films with varying DE layers. These are a nominally DE-free ('HZO only') film, a film with 1.5\,nm Al$_2$O$_3$ and a film with 2\,nm Al$_2$O$_3$. Finally, pulse measurements were performed at temperatures from 100-300\,K. Assuming that the tunneling mechanism and thus the charge injection in the Al$_2$O$_3$ layers is temperature-dependent \cite{hsiang2021bilayer}, modifying the sample temperature gives control over the trap occupation, thereby also modifying E$_{trap}$. 

\subsection{Measurement setup}
Room temperature measurements were performed on a Cascade Microtech probe station. Pulses were applied by a Keithley 4225 Pulse Measurement Unit (PMU) controlled with a Keithley 4200A-SCS parameter analyzer. The lower measurement range is increased by using 4225 Remote Preamplifier/switch Modules (RPMs).

Temperature dependent measurements were conducted on a Lake Shore Cryogenic CPX-VF probe station, and liquid nitrogen was used as a coolant. This could be used to cool the system down to 80\,K, but we limited our lower temperature to 100\,K to ensure stable temperature operation. Kapton tape was placed between the sample and stage to maintain electrical isolation from the grounded cryo stage. Pulses were applied from the same PMUs as used in the room temperature measurements. 

\section{Results}
\subsection{Polarization loss as a function of delay time}\label{PlossTime}

\begin{figure}[h!]
\begin{center}
\includegraphics[width=15cm]{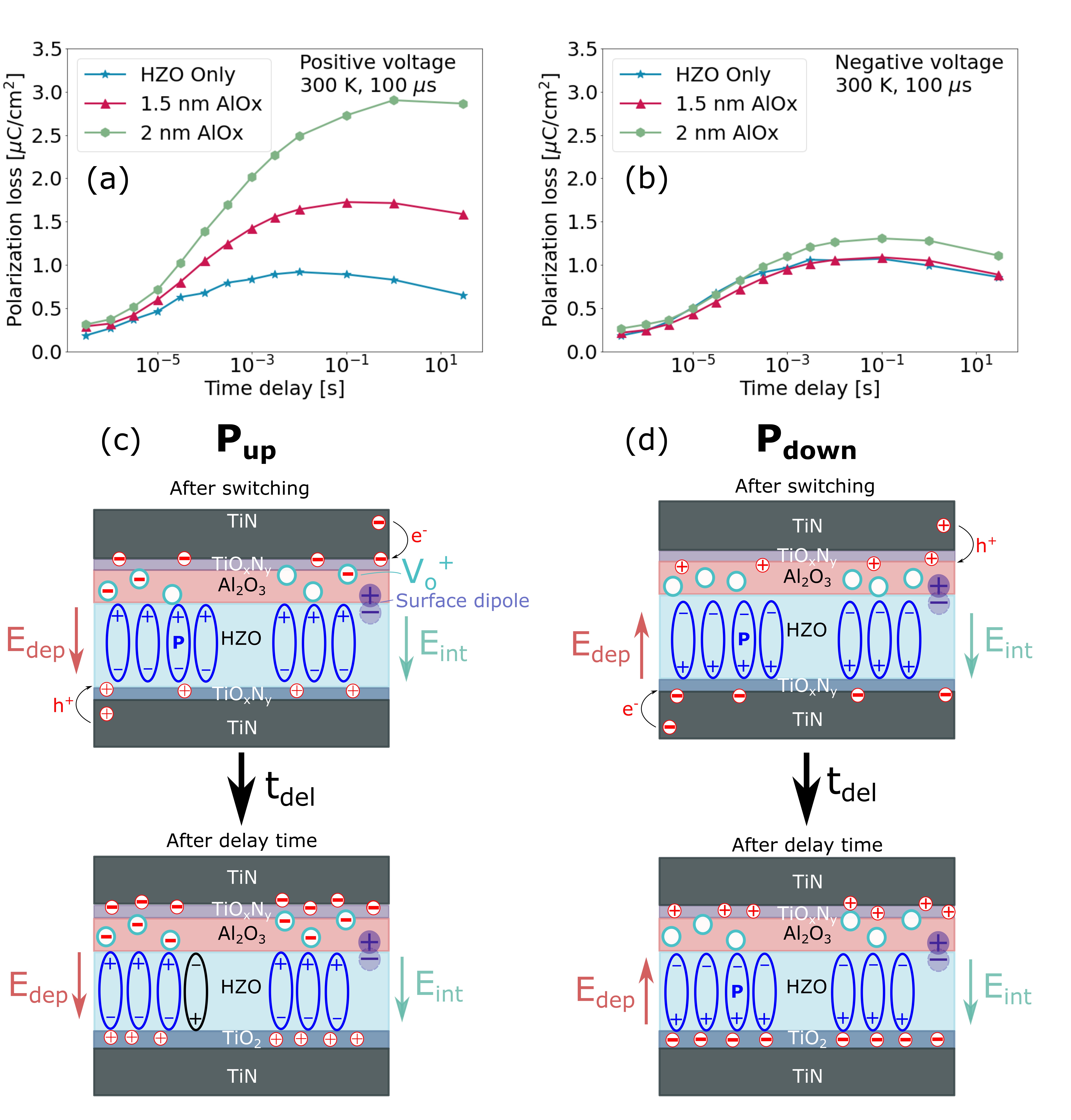}
\end{center}
\caption{Polarization loss (P$_{loss}$) as a function of the delay time under no external bias (t$_{del}$) for (a) positive polarity (P$_{up}$ state) and  (b) negative polarity (P$_{down}$ state). Voltage amplitudes are 3.5\,V (HZO only), 5.5\,V (1.5\,nm Al$_2$O$_3$), 6\,V (2\,nm Al$_2$O$_3$). Schematic depictions of the movement of charges within the stack for (c) the P$_{up}$ state and (d) the P$_{down}$ state, during the delay time.} \label{fig:3} 
\end{figure}

First, devices were woken-up at room temperature by applying 1000 electric field cycles at a frequency of 100\,kHz. Then, the polarization loss was measured as a function of the delay time at 0\,V bias. Figure \ref{fig:3} shows this relationship for devices of each oxide thickness, for pulse widths of 100\,$\mu$s, in the P$_{up}$ and P$_{down}$ states. 

One significant trend shows that the polarization loss at a given time delay, for P$_{up}$ (positive voltages), increases as the thickness of the dielectric increases. The polarization loss difference between films becomes even stronger with increasing delay time until saturating after roughly 100\,ms. The greater polarization loss can be attributed to the increased depolarization field stemming from a thicker non-switching layer, combined with less effective charge injection at the top HZO interface due to the thick DE layer. After time, the polarization loss in a given polarity is seen to saturate. The longest time delay used in our experiments was 30\,s; thus, this saturation implies that on the time scales investigated here, we see a saturating short-term retention loss effect, which is further analyzed in the next section.

Finally, there is a clear asymmetry in the magnitude of P$_{loss}$ measured in each polarity. This asymmetry is larger for thicker Al$_2$O$_3$ layers due to the increasing depolarization field, which requires a larger amount of trapped charge to produce a compensating field E$_{trap}$. At the same time, charge injection is hindered with a thicker Al$_2$O$_3$ layer. The asymmetry of the polarization loss highlights the role of E$_{int}$ in destabilizing the polarization predominantly in one direction. The aggregation of positively charged oxygen vacancies at the top Al$_2$O$_3$ interface is consistent with a quasi-static internal bias field that points down. As shown schematically in figure \ref{fig:3}c, the internal bias field is assumed to come from one or more effects. An intrinsic effect to the bilayer material system is the difference in areal oxygen density in Al$_2$O$_3$ and HZO, which should lead to a surface dipole. Negatively charged oxygen ions at the interface will preferentially move to the material with a lower areal oxygen density, in our case HZO, forming an internal field pointing in the opposite direction compared to the dipole. As in the case of HZO only films, oxygen vacancies may be generated near the top of the material stack \cite{kita2009origin}. Finally, the physical separation interposed by a thicker dielectric layer at one electrode interface and a more ideal metal/ferroelectric opposing interface may cause the ferroelectric domains to polarize in the direction of the high concentration of free electrons at the more ideal metal electrode interface \cite{lomenzo2015tan}. Besides directly acting against the polarization in the P$_{up}$ state, E$_{int}$ can also inhibit charge injection which would stabilize the state. Conversely, charge migration is facilitated in the direction of the internal field (figure \ref{fig:3}d), leading to a lower polarization loss in the P$_{down}$ state. This effect has direct implications for device operation and can be observed for example in the retention characteristics of bilayer FTJs, where retention loss acts preferentially on one state \cite{max2019retention}. This will be discussed in more detail in section \ref{PlossTemp}. 

\subsection{Polarization loss as a function of pulse width}\label{PlossPW}
\begin{figure}[h!]
\begin{center}
\includegraphics[width=15cm]{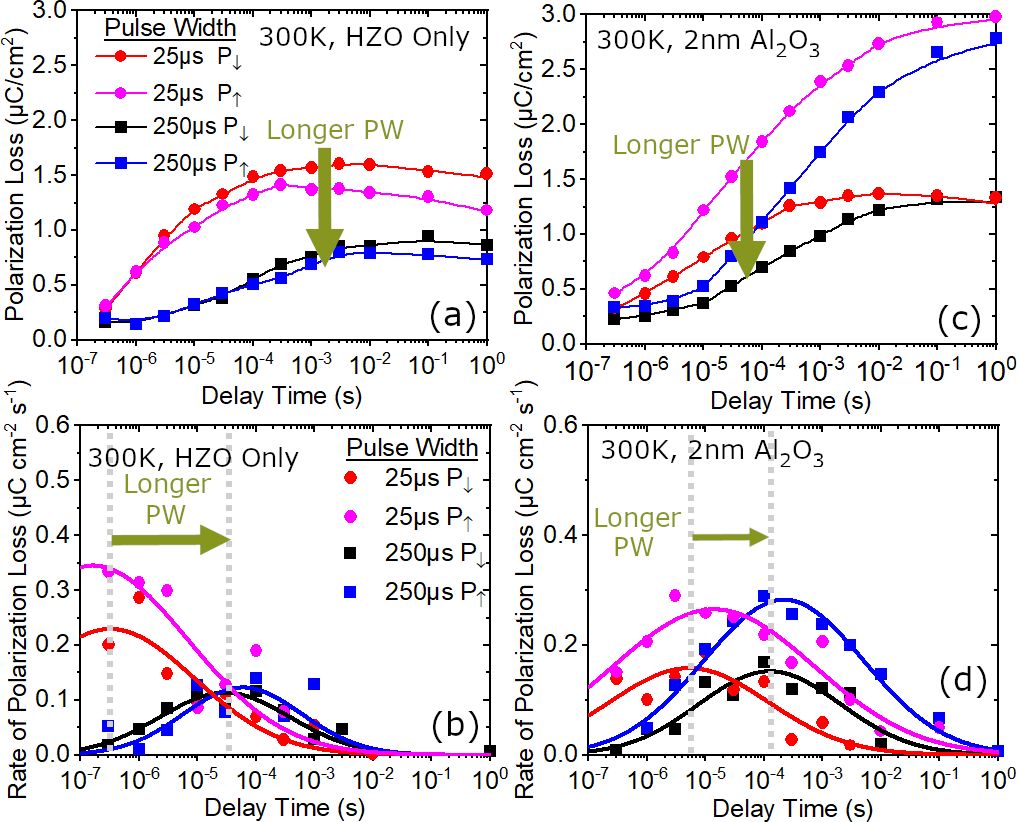}
\end{center}
\caption{Polarization loss vs. delay time at 300\,K for the (a) HZO only film and (c) the HZO stack with a 2 nm Al$_2$O$_3$ top interfacial layer. Rate of polarization loss vs. delay time at 300\,K for the (b) HZO only film and (d) the HZO stack with a 2 nm Al$_2$O$_3$ top interfacial layer. The data shows up and down polarization states, as well as 25\,$\mu$s and 250\,$\mu$s applied pulse widths}\label{fig:4}
\end{figure}

The quantity of the injected charge carriers that screen the depolarization field should be modulated by the pulse width of the applied bias. Since the injected charge carriers are predicted to reduce the depolarization field, it is expected that a resulting decrease in the polarization loss with increasing pulse width would adhere to the charge injection screening model. Moreover, since longer pulse widths inject more screening charges, the lower magnitude of the depolarization field should also lead to a slower rate of the polarization loss. This can be understood by a switching kinetics model, whereby the depolarization field and the delay time take the place of the applied bias and time as a function of switched polarization in conventional switching kinetic models \cite{zhukov2010dynamics, materano2020polarization}. Thus, when the depolarization field is stronger due to a lower quantity of screening charge, the ‘back-switching’ is faster and more charge is back-switched, similar to the case of a switching kinetic measurement where an applied voltage bias increases the amount of switched charge within a given timescale. 

Figure \ref{fig:4}a shows the polarization loss at 300\,K of the HZO only film as a function of delay time for the up and down polarization states after the longest (250\,$\mu$s) and shortest (25\,$\mu$s) pulse widths deployed in this investigation. The 250\,$\mu$s pulse width reduced the total polarization loss in both up and down polarization states compared to the 25\,$\mu$s applied voltage pulse width. The reduction in transient polarization loss with increasing pulse width is consistent with enhanced charge injection and subsequent screening of the ferroelectric charge. Both up and down polarization states have similar magnitudes of polarization loss after 1\,s delay time independent of pulse width. This can be understood as the film reaching a quasi-state of equilibrium within this timescale, whereby the depolarization field effectively neutralizes itself by causing charge injection that enhances screening of the ferroelectric charge, causing the polarization loss to level off and saturate after more than 100\,ms of delay time. 

Not only does the magnitude of the polarization loss change with the pulse width, but there is a strong shift in the distribution of the polarization relaxation processes (‘back-switching’) in time, as shown in figure \ref{fig:4}b. The rate of the polarization loss is obtained by taking the derivative of the polarization loss with respect to the delay time. A Gaussian distribution is used to fit the experimental data, indicating the stochastic nature of the polarization loss process. The unit of the rate of polarization loss is consistent with its representation as a current density that can be understood as the ‘back-switching’ or polarization relaxation current. Independent of the polarization state, longer pulse widths decrease the rate of polarization loss and shift the polarization loss to longer times. Both the decrease in magnitude and the delay of the polarization loss is consistent with a lower depolarization field, which can be explained by enhanced screening caused by injected charge carriers. The slight differences in the polarization relaxation current peak for the two polarization states may be due to a small pre-existing internal bias field, as well as an asymmetric charge injection process at the top and bottom TiN/FE interfaces, attributable to the different degrees of oxidation and chemical properties of the top and bottom TiN interfaces \cite{baumgarten2021impact}.

The film stack with a 2\,nm Al$_2$O$_3$ top interfacial layer exhibits a difference with regard to the polarization loss dependency on pulse width at 300\,K,  Figure \ref{fig:4}c,  although longer pulse widths still reduce the magnitude of polarization loss for a given delay time. However, the Al$_2$O$_3$ top interface layer causes a more than 2x increase in polarization loss in the P$_{up}$ compared to the P$_{down}$ state. The difference in the polarization loss between the up and down polarization states can no longer be compensated by the 250\,$\mu$s pulse width, unlike the HZO only film where both polarization states had a similar loss with the longer pulse width. Additionally, the loss of the polarization in the up state does not saturate after a 1\,s delay time, unlike the down state. The significantly increased polarization loss in only P$_{up}$ with Al$_2$O$_3$ and the inability of the longer pulse-width to fully compensate the polarization loss between the two polarization states again indicates that a significant internal bias field, pointing downward, exists in the HZO film, caused by the Al$_2$O$_3$ film. Such an internal bias field is expected to be independent of the depolarization field since a large increase in the amount of injected charge carriers would be expected to neutralize the depolarization field and equalize the stability of each polarization state with increasing pulse width. 

Compared to the HZO only film stack, the introduction of the 2\,nm Al$_2$O$_3$ top interface layer causes a smaller shift of the polarization loss current peak with increasing pulse width, Figure \ref{fig:4}d. This smaller shift with pulse width may be due to the compromised ability of injected carriers to sufficiently screen the depolarization field, resulting in only a weak modulation of the characteristic ‘back-switching’ time with pulse width. However, the fact that some delay is observed at longer pulse widths suggests that injected carriers still influence the overall dynamics of the polarization loss process in the Al$_2$O$_3$ stack at room temperature for both polarization states. Considering that the polarization loss current is wider and peaks at longer delay times with the 2\,nm Al$_2$O$_3$ interface layer stack compared to the HZO only stack, charge carrier injection and screening under the depolarization field could be inhibited by the dielectric layer since it proceeds at a slower and wider ranging pace. The biggest difference when the 2\,nm Al$_2$O$_3$ is incorporated into the top interface of the HZO stack is in the sustained higher magnitude of the rate of polarization loss that the P$_{up}$ state undergoes, indicating that the built-in electric field causes a significant increase in the amount of ‘back-switched’ charge in this stack structure compared to the HZO only film stack. 

\subsection{Polarization loss as a function of temperature}\label{PlossTemp}
\begin{figure}[h!]
\begin{center}
\includegraphics[width=15cm]{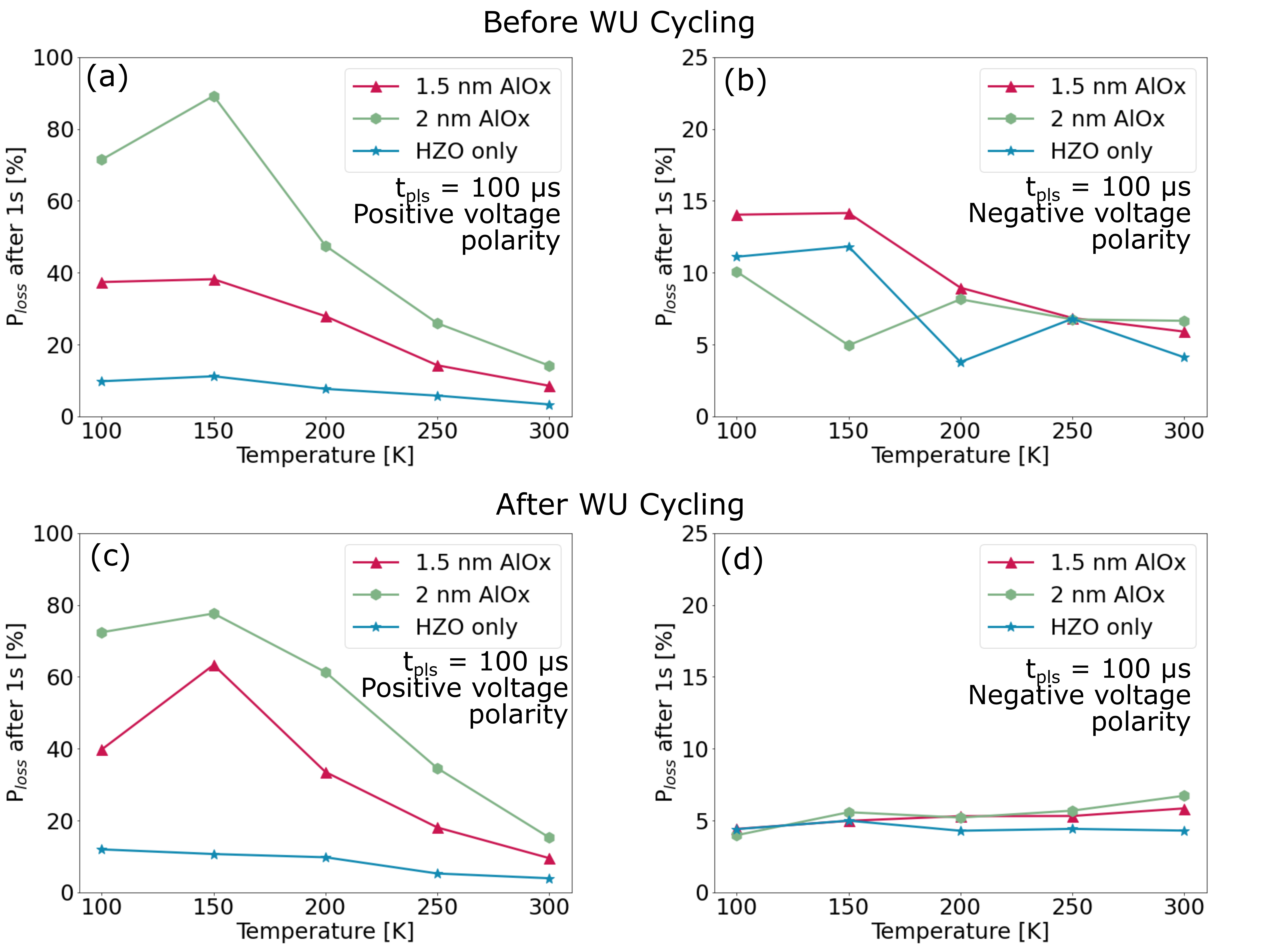}
\end{center}
\caption{Polarization loss (P$_{loss}$) as a function of temperature, measured after 1\,s delay time under no external bias (t$_{del}$ = 1\,s), before wake-up cycling for (a) positive voltage polarity (P$_{up}$ state) and  (b) negative voltage polarity (P$_{down}$ state); after wake-up cycling for (c) positive voltage polarity (P$_{up}$ state) and  (d) negative voltage polarity (P$_{down}$ state). In all cases the polarization loss is normalized to the total switched charge measured on the initial pulse. Voltage amplitudes are 3.5\,V (HZO only), 5.5\,V (1.5\,nm Al$_2$O$_3$), 6\,V (2\,nm Al$_2$O$_3$).}\label{fig:5}
\end{figure}

Finally, devices were investigated as a function of temperature. Both cycled (woken-up) and uncycled devices were measured at 50\,K intervals from 100-300\,K. Wake-up was always performed at room temperature, as it is known to have a strong temperature dependence due to the energy barrier for the redistribution of charge traps \cite{starschich2016evidence}, which would therefore influence our polarization loss measurements. As above, wake-up was performed for 1000 cycles at 100\,kHz. To compare between measurements, P$_{loss}$ was always extracted after a delay time of 1\,s. There is both an Al$_2$O$_3$ thickness dependence and a temperature dependence on the total switchable charge 2Pr; as such, the data plotted here are normalized following equation \ref{eqn:2}.

Figures \ref{fig:5}a \& \ref{fig:5}b demonstrate a strong asymmetry in the temperature-dependent behaviour of P$_{loss}$ when Al$_2$O$_3$ is introduced into the device stack. Contrary to the room-temperature dependent data shown in \ref{fig:1}, we also see a slight asymmetry (around 5\%) in the behaviour of the nominally HZO only film. As shown in that figure, the polarization loss is larger for positive voltages (P$_{up}$, electrons injected at the TE) than negative voltages (P$_{down}$, electrons injected at the BE). Further, a strong dependence on Al$_2$O$_3$ thickness is observed in for P$_{up}$ while P$_{down}$ shows no such thickness dependence on polarization loss. This offers further evidence that the dominating factor in P$_{loss}$ is the interplay between electron injection and the internal bias field. Assuming that electrons are the primary charge carrier being injected into the interface states, P$_{loss}$ in the negative polarity would be mainly determined by the properties of the interfacial layer (IL) at the bottom electrode, which is nominally identical for all three films.

From this we can conclude that all three films have asymmetrical ILs, although the asymmetry is relatively small in the case of the nominally HZO-only film compared to the FE/DE bilayers. To further understand this, we can consider the process flow of the stacks. Firstly, a vacuum break occurs between the TiN and HZO depositions. Additionally, the bottom TiN is oxidized during the first few ALD cycles. This forms a barrier for oxygen scavenging which occurs during the crystallization anneal \cite{szyjka2020enhanced}. On the other hand, the top HZO/TiN or HZO/Al$_2$O$_3$ interface is more reactive during the anneal \cite{fengler2018relationship}. As such, a large amount of oxygen vacancies are expected to form near the TE, which dominate the switching properties of the film prior to electric field cycling. In fact, the formation of a tetragonal IL near the top electrode has been observed \cite{grimley2016structural} and attributed to this oxygen scavenging effect \cite{goh2020oxygen, mittmann2021impact}. 

In figures \ref{fig:5}c \& \ref{fig:5}d, the same experiments were repeated on woken-up films. Strikingly, there is no change in the P$_{loss}$ observed for positive voltages, while the temperature dependence of the polarization loss for negative voltages is reduced. In fact, as is shown in supplementary figure S2, the temperature dependence for negative voltages changes sign, with a lower P$_{loss}$ at lower temperatures. 

Comparing the data before and after wake-up cycling, we can conclude that the depolarization field originating from the TE IL remains unchanged with wake-up cycling. At the same time, the temperature dependence of P$_{loss}$ for negative voltages indicates that the polarization state can be better stabilized after wake-up. As discussed above, the internal bias field E$_{int}$ plays a large role in both destabilizing the P$_{up}$ state and also in suppressing charge injection during the delay time. As the temperature is decreased, fewer charges are available to act against the internal field. This strengthens the internal field and thereby increasingly favours P$_{down}$ over P$_{up}$.  Thus, the electrical properties of the films are dominated by the asymmetry in charge trapping at the two interfaces, as has been previously demonstrated for MIM \cite{weinreich2009impact}, MFS \cite{zacharaki2020reliability} and MFM capacitors \cite{pesic2016impact} and remains one of the major reliability issues for ferroelectric thin films.

\subsection{Consequences for ferroelectric memory devices}\label{Results_FeMEM}
The results demonstrated here have an impact on the design and operation of ferroelectric-based memory cells, such as FeFET and FTJ. On the design side, the results indicate that there is a trade-off in device speed and compensation of the polarization charges. As previously shown in literature \cite{fontanini2022interplay}, having a larger density of interface traps helps to stabilize a higher 2Pr value in the case of a thicker interfacial oxide (for example, in bilayer FTJ devices). Our results indicate that this leads to a certain `settling time' during which a charging current can be observed, on a time scale which does not vary greatly with the applied pulse parameters (figure \ref{fig:4}d). While exploring a larger parameter space may yet yield a similar difference in the rate of polarization loss as seen for HZO films (figure \ref{fig:4}b), it is also plausible that the slow charge trapping is an inevitable result of the intrinsic contributions of the internal bias field and the depolarization field. Of these two, the internal bias field in particular could be further researched in order to better stabilize the binary switching behavior. This may be achieved through interfacial engineering and modifying the oxygen gradient in the film stack, in particular. The highly asymmetric behavior demonstrated in figures \ref{fig:3}, \ref{fig:4}c,d \& \ref{fig:5} further strengthens the hypothesis that these short time dynamics are strongly impacted by a unidirectional internal bias field.

The polarization loss shown in figures \ref{fig:4}b,d can be translated into a current, which needs to be accounted for in the design of the readout circuitry especially in the case of FTJ devices that typically feature a comparatively low read current. Accordingly, the minimum read latency of such memory devices is limited to the time after the peak of the polarization loss. In the case of FeFET, the characterization described here may also aid in both design optimization and determining suitable write operation parameters in order to reduce read latency, with potential trade-offs in short-term retention.

\section{Conclusions \& Outlook}

We have demonstrated a novel characterization method for analyzing the backswitched charge in ferroelectric stacks on small time scales. This is a complementary method to standard retention measurements which are used to investigate device reliability in ferroelectric HZO or bilayer stacks \cite{mueller2012reliability, max2019retention}, in that it is used to investigate the fast dynamics of the system after switching. The measurement gives important insights into how polarization is stabilized in ferroelectric films, particularly when integrated with an intentional dielectric layer. A rapid polarization loss is seen in the first 100\,ms after switching, which is a critical time frame for applications of bilayer FTJs \cite{covi2021ferroelectric} or FeFETs \cite{ni2018critical}. Indeed, the current arising from backswitching which can be extracted from the rate of polarization loss in figure \ref{fig:4} is critical when compared to the read current of an FTJ device. 

By analyzing the polarization loss as a function of delay time, pulse width and temperature, we propose a charge trapping and screening model to explain the fast time dynamics of ferroelectric layers. To a lesser extent, this effect is also present in single-layer HZO films. Charge trapping, driven by the depolarization and internal bias fields, occurs in the first tens of ms after switching and helps to stabilize switched polarization by screening the polarization charges. Simultaneously, the depolarization field leads to back-switching of some domains. These competing effects are evident in a peak in the rate of polarization loss which for HZO films is highly dependent on the parameters of the applied pulses. For samples with a dielectric interlayer, the rate of polarization loss is less dependent on the pulse parameters for the range of pulse widths explored here, as the dynamics are mainly controlled by the larger internal bias field and longer pulses may be needed to inject a greater quantity of screening charge across the Al$_2$O$_3$ layer. 

Beyond the scope of the analysis here, the data produced by this measurement could be applied to analyze imprint effects or to better identify the time dynamics of charge injection in HZO and HZO/DE bilayers. The investigation of the dynamics of fast polarization loss in ferroelectric films offers additional material properties which should be carefully designed for, in order to achieve high-performing ferroelectric devices. 





\section*{Conflict of Interest Statement}

The authors declare that the research was conducted in the absence of any commercial or financial relationships that could be construed as a potential conflict of interest.

\section*{Author Contributions}
SL conceived the experiments, fabricated the samples, analyzed data and wrote the manuscript. PL contributed to data analysis, including development of the analytical model, and provided supervision. ME performed the measurements and contributed to analysis and manuscript writing. BX performed the measurements. TM, US and SS provided supervision, funding, and input on the manuscript. 


\section*{Funding}
\section*{Acknowledgments}
SL and SS were financially supported by the European Union through the BeFerroSynaptic project (GA:871737), by the DFG through the FLAG-ERA JTC 2019 grant SOgraphMEM (MI 1247/18-1) and by the DFG through the Memristec priority program, in the scope of the project ReLoFemRis (SL 305/2-1). B.X. and P.D.L. were financially supported by the Deutsche Forschungs Gemeinschaft DFG within the following projects (Zeppelin (433647091) and Homer (430054035)). TM and US were financially supported out of the Saxonian State budget approved by the delegates of the Saxon State Parliament.

\section*{Supplemental Data}
Supplemental data can be found at ...

\section*{Data Availability Statement}
The datasets generated for this study are available from the corresponding author upon reasonable request.

\bibliographystyle{plain} 
\bibliography{bibliography}




\end{document}


\onecolumn

\maketitle

\section{Measurement on the falling edge of pulses}

\begin{figure}[h!]
\begin{center}
\includegraphics[width=15cm]{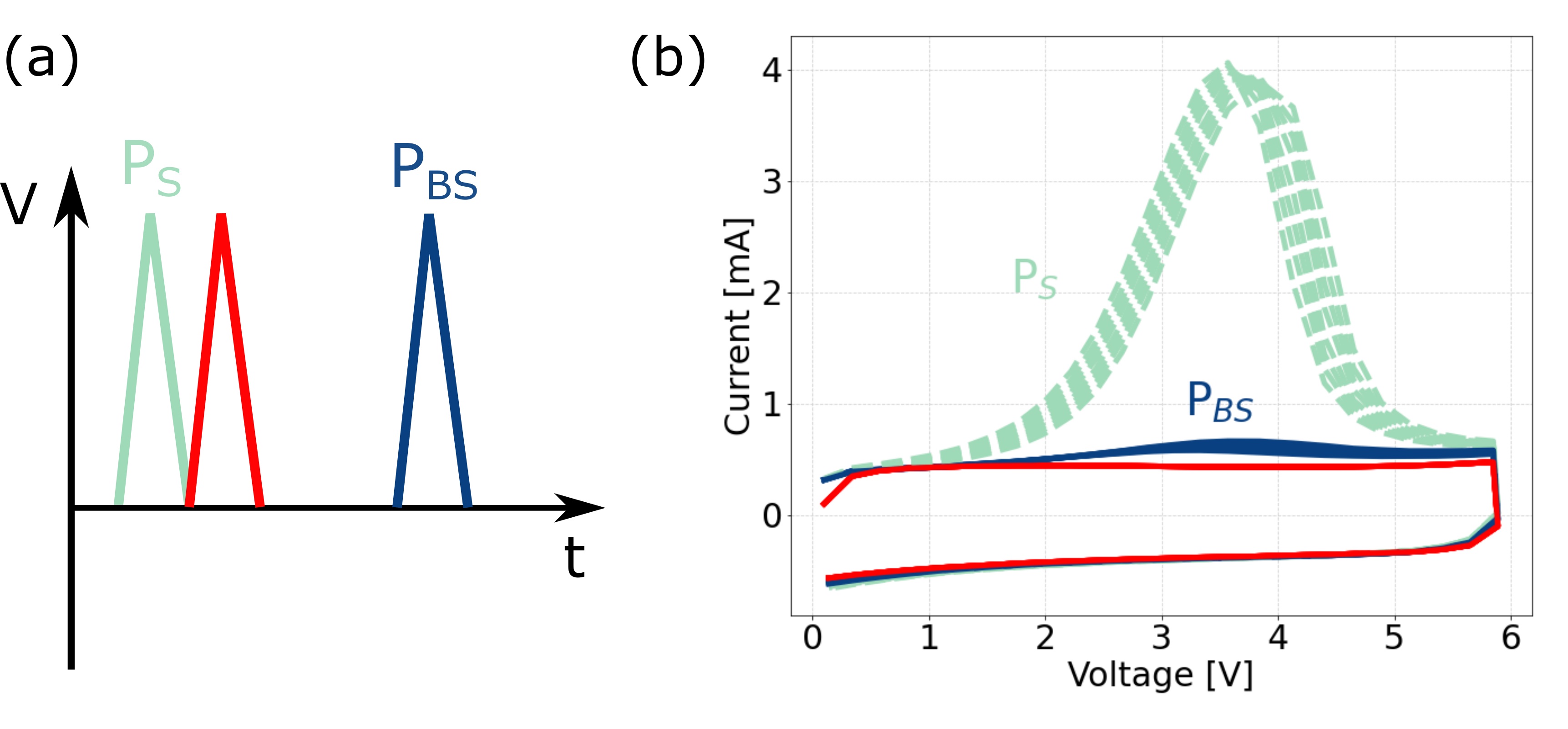}
\end{center}
\caption{Measurement of P$_S$ and P$_{BS}$ including falling edge.}\label{fig:S1}
\end{figure}

In preliminary experiments, a modified pulse train was applied as in figure \ref{fig:S1}a. Figure \ref{fig:S1}b depicts the P$_S$ and P$_{BS}$ pulses where the current has been measured on both the rising and falling edges of the pulses. The pulse in red is the 'background' pulse (BG), measured 20\,ns after P$_S$, which was subtracted from both P$_S$ and P$_{BS}$. For I \textless{} 0\,mA (corresponding to the falling pulse edge), we see that all three pulses perfectly overlap. Therefore, they are cancelled in the integrals of equations 1 and 2. Since the total number of measurement points for the whole measurement (a single measurement is one set of pulse parameters, measured at all time delays t$_{del}$) is limited by the setup, we used the above conclusion to measure on the rising edges only. This allowed us to maximise the number of time delays we could investigate without introducing additional unwanted delays into the measurement. 

\section{Temperature dependence of polarization loss in negative polarity}

\begin{figure}[h!]
\begin{center}
\includegraphics[width=15cm]{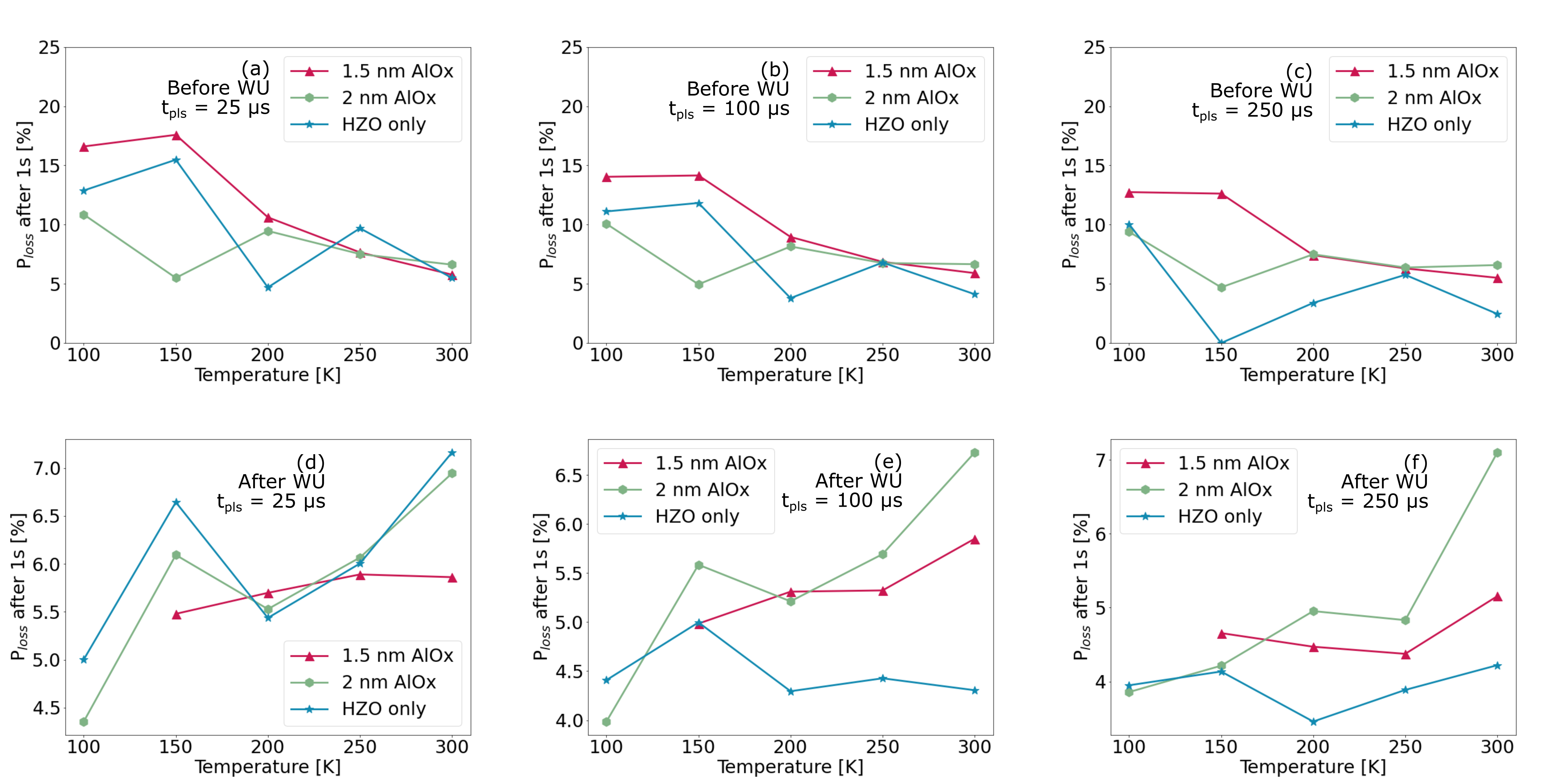}
\end{center}
\caption{Polarization loss in negative polarity, as a function of temperature, for pulse widths of (a) 25\,$\mu$s, (b) 100\,$\mu$s and (c) 250\,$\mu$s (uncycled films); (d) 25\,$\mu$s, (e) 100\,$\mu$s and (f) 250\,$\mu$s (woken-up films)}\label{fig:S2}
\end{figure}

Figure \ref{fig:S2} shows the temperature dependence of polarization loss in the negative voltage polarity, for both uncycled (a-c) and woken-up (d-f) films. Before wake-up, the polarization loss increases with reducing temperature, which is consistent with the charge trapping and screening model presented in the main text. After wakeup cycling, the temperature dependence of P$_{loss}$ is greatly reduced, even tending towards the inverse behaviour. With wakeup cycling, we expect a redistribution in charged vacancies incorporated into the stack during processing. Nonetheless, the internal field arising from the asymmetry of the electrodes persists. Thus at lower temperatures, where there are fewer mobile charges to screen this internal field, we observe a strong favoring of one polarization state over the other. Nonetheless, the dependence of P$_{loss}$ on pulse width (compare figures \ref{fig:S2}(d-e)) still indicates that we have an influence of charge trapping on the polarization state.